\documentclass[12pt,a4paper]{iopart}
\usepackage{iopams}
\usepackage{graphicx}
\usepackage{epsf}
\usepackage{epsfig}
\usepackage[usenames]{color}
\usepackage{amssymb}
\usepackage{times}
\usepackage{hyperref}

\newcommand{\ie}{\textit{i.e.}~}
\newcommand{\eg}{\textit{e.g.}~}
\newcommand{\vb}{v_\mathrm{b}/c}
\newcommand{\eqref}[1]{(\ref{#1})}
\begin{document}

\title[Black-hole production from ultrarelativistic collisions]
{Black-hole production from ultrarelativistic collisions}

\author[L. Rezzolla and K. Takami]
{
Luciano Rezzolla$^{1,2}$
and 
Kentaro Takami$^{1}$ 
}

\address{$^{1}$Max-Planck-Institut f{\"u}r Gravitationsphysik, Albert
Einstein Institut, Golm, Germany}
\address{$^{2}$Department of Physics, Louisiana State University, Baton Rouge, LA USA}

\begin{abstract}
Determining the conditions under which a black hole can be produced is a
long-standing and fundamental problem in general relativity. We use
numerical simulations of colliding selfgravitating fluid objects to study
the conditions of black-hole formation when the objects are boosted to
ultrarelativistic speeds. Expanding on previous work, we show that the
collision is characterized by a type-I critical behaviour, with a black
hole being produced for masses above a critical value, $M_{\rm c}$, and a
partially bound object for masses below the critical one. More
importantly, we show for the first time that the critical mass varies
with the initial effective Lorentz factor $\langle \gamma \rangle$
following a simple scaling of the type $M_{\rm c} \sim K \langle \gamma
\rangle^{-1.0}$, thus indicating that a black hole of infinitesimal mass
is produced in the limit of a diverging Lorentz factor. Furthermore,
because a scaling is present also in terms of the initial stellar
compactness, we provide a condition for black-hole formation in the
spirit of the hoop conjecture.
\end{abstract}



\section{Introduction}

There is little doubt that black-hole formation represents one of the
most intriguing and fascinating predictions of classical general
relativity. There is abundant astronomical evidence that black holes
exist, and a number of considerations supporting the idea that under
suitable conditions gravitational collapse is
inevitable~\cite{Wald84}. In addition, there is overwhelming numerical
evidence that black-hole formation does take place in a variety of
environments~\cite{Rezzolla:2011}. Yet, a rigorous definition of the
sufficient conditions for black-hole formation is still lacking. Hence,
it is not possible to predict whether the collision of two compact
objects, either stars or elementary particles, will lead to the formation
of a black hole.

However, we are not without ideas and the intuitive \textit{hoop
  conjecture} proposed by Thorne in the '70s, provides us with some
reasonable guidelines~\cite{Thorne72a}. We recall that the conjecture
states that a black hole is formed if an amount of ``mass-energy'' $E$
can be compressed to fit within a hoop with radius equal or smaller than
the corresponding Schwarzschild radius, \ie if \hbox{$R_{\mathrm{hoop}}
  \leq R_\mathrm{s} = 2GE/c^4$}, where $G$ is gravitational constant and
$c$ the speed of light. Even though it can be made precise under
particular circumstances~\cite{senovilla_2008_rhc}, the hoop conjecture
is not meant to be a precise mathematical statement and, in fact, it is
difficult to predict if the above-mentioned collision will compress
matter sufficiently to fit within the limiting hoop. Loosely speaking,
what is difficult is to determine which part of the ``kinetic energy'' of
the system can be accounted to fit within the hoop. Since at the
collision the conversion of kinetic energy into internal energy is a
highly nonlinear process, any quantitative prediction becomes rapidly
inaccurate as the speeds involved approach that of light.

As stated above, the hoop conjecture is purely classical. A
quantum-mechanical equivalent is not difficult to formulate, although not
very stringent, as it simply implies that a black hole will be formed at
Planck-energy scales. The predicting power does not improve significantly
when considering the conditions of black-hole formation in
higher-dimensional theories of gravity (see, \eg
\cite{Argyres_1998,Yoshino_2003,Yoo_2010}). In these frameworks, the
energy required for black-hole formation might be significantly
smaller~\cite{Argyres_1998}, thus providing the possibility of producing
them in the Large Hadron Collider (LHC)~\cite{Dimopoulos_2001}, but no
firm conclusion has been reached yet.

Clearly, although numerical simulations represent a realistic route to
shed some light on this issue (see, \eg,
\cite{Sperhake:2008ga,Shibata2008,Sperhake2009}), even the simplest
scenario of the collision of two compact objects at ultrarelativistic
speeds is far from being simple and it is actually very challenging. A
first step was taken by Eardley and Giddings~\cite{Eardley_2002}, who
have studied the formation of a black hole from the head-on collision of
two plane-fronted gravitational waves with nonzero impact parameter
(previous work of D'Eath and
Payne~\cite{Death1992a,Death1992b,Death1992c} using different methods had
considered a zero impact parameter). In all of these analyses each
incoming particle is modelled as a point particle accompanied by a
plane-fronted gravitational shock wave corresponding to the
Lorentz-contracted longitudinal gravitational field of the particle. At
the instant of collision the two shock waves pass through one another and
interact through a nonlinear focusing and shearing. As a result of their
investigation, a lower bound was set on the cross-section for black hole
production, \ie $\sigma > 32.5(GE/2c^4)^2$, where $E$ is the
centre-of-mass (lab) energy. More recently, and in a framework which is
closer to the one considered here, this problem has been investigated by
Choptuik and Pretorius~\cite{Choptuik:2010a}, who studied the collision
of two classical spherical solitons, with a total energy of the system in
the lab frame $E=2 \gamma_{\mathrm{b}} m_0 c^2$, where $m_0$ is the
``rest-mass'', $\gamma_{\mathrm{b}} \equiv 1/\sqrt{1-
  v^2_{\mathrm{b}}/c^2}$ and $v_{\mathrm{b}}$ the boost velocity. They
were then able to show that for collisions with sufficiently high boost,
\ie $\gamma_{\mathrm b} \gtrsim 2.9$, a black hole can be formed.

In this work we report for the first time on black-hole production from
the collision of two compact, self-gravitating, fluid objects boosted at
ultrarelativistic speeds\footnote{While this work was being reviewed, a
  similar investigation by East and Pretorius has also been
  reported~\cite{East2012}.}. There are several important differences
with the previous investigations
in~\cite{Death1992a,Death1992b,Death1992c,Eardley_2002,Choptuik:2010a}. Differently
from~\cite{Death1992a,Death1992b,Death1992c,Eardley_2002}, in fact, our
colliding objects are not in vacuum and are not treated as point
particles. Rather, our relativistic stars are obviously extended and
self-gravitating objects, thus with a behaviour that is intrinsically
different. Also, differently from~\cite{Choptuik:2010a}, our objects are
not described as scalar fields, but as fluids and thus represent a more
realistic description of baryonic matter, such as the one employed when
simulating relativistic heavy-ion collisions~\cite{Rischke1995}. These
intrinsic differences also make the comparison with the works
of~\cite{Death1992a,Death1992b,Death1992c,Eardley_2002} very hard if
possible at all. On the other hand, many analogies exist with the
collision of bosons stars considered in~\cite{Choptuik:2010a}, and that,
as we will discuss below, can be interpreted within the more general
description of black-hole production from ultrarelativistic collisions
that we provide in this work.

Given these basic differences, it is not surprising that our results also
show qualitative differences with respect to previous investigations. The
most important of these differences is that we find that a black hole can
be produced even from zero initial velocities if the initial masses are
large enough; this behaviour is clearly absent in all previous results,
where instead a critical initial boost is
necessary~\cite{Death1992a,Death1992b, Death1992c, Eardley_2002,
  Choptuik:2010a}. In addition, we show that for each value of the
effective Lorentz factor, $\langle \gamma \rangle$, a critical initial
mass exists, $M_{\rm c}$, above which a black hole is formed and below
which matter, at least in part, selfgravitates. More importantly, both
$M_{\rm c}$ follows a simple scaling with $\langle \gamma \rangle$.

\begin{table}
\caption{Initial properties of the isolated stars leading to
  supercritical models nearer to the critical line. Reported in the
  various columns are the central density $\rho_c$, the gravitational
  mass $M$, the rest mass $M_0$, the radius $R$, the compactness $M/R$,
  the initial velocity and Lorentz factor $v_b$ and $\gamma_b$, and the
  corresponding effective value $\langle \gamma\rangle$ in units where
  $c=1=M_{\odot}$.}
\label{table:ID} 
\begin{indented}
\lineup
\item[]\begin{tabular}{@{}*{8}{c}}
\br
$\rho_\mathrm{c}\times 10^{4}$ & $M\times 10$ & $M_{0}\times 10$ 
& $R\times 10^{-1}$ & $(M/R)\times 10^{2}$ 
& $v_{\mathrm{b}}$ & $\gamma_{\mathrm{b}}$ & $\langle \gamma \rangle$\cr
\mr
5.31613  &  8.970  & 9.348 &  1.101  &  8.148  &  0.30  &  1.048  &  1.033\cr
4.90000  &  8.501  & 8.837 &  1.111  &  7.653  &  0.50  &  1.155  &  1.107\cr
3.75000  &  7.043  & 7.267 &  1.140  &  6.179  &  0.70  &  1.400  &  1.284\cr
3.00000  &  5.947  & 6.103 &  1.160  &  5.126  &  0.80  &  1.667  &  1.497\cr
2.50000  &  5.142  & 5.257 &  1.174  &  4.379  &  0.86  &  1.960  &  1.757\cr
2.05000  &  4.362  & 4.443 &  1.187  &  3.674  &  0.90  &  2.294  &  2.116\cr
1.20000  &  2.728  & 2.759 &  1.213  &  2.248  &  0.95  &  3.202  &  3.618\cr
\br
\end{tabular}
\end{indented}
\end{table}

\section{Methodology}

The numerical setup employed in our simulations is the \emph{same}
presented extensively in our previous
works~\cite{Kellermann:08a,Kellermann:10,Radice:10}, and the interested
reader can find in these references all the needed technical details. It
is sufficient to remind here that we use an axisymmetric code to solve in
two spatial dimensions, $(x,z)$, the set of the Einstein and of the
relativistic-hydrodynamic equations~\cite{Kellermann:08a}. The
axisymmetry of the spacetime is imposed exploiting the ``cartoon''
technique, while the hydrodynamics equations are written explicitly in
cylindrical coordinates. All the simulations use an ideal-gas EOS,
$p=(\Gamma-1)\rho\epsilon$, where $\rho$ is the rest-mass density,
$\epsilon$ the specific internal energy, and $\Gamma=2$. The initial
configurations consist of spherical stars, constructed as
in~\cite{Kellermann:10,Radice:10} after specifying the central density,
$\rho_c$, where the latter also serves as parameter to determine the
critical model. The stars have an initial separation $D$ and are boosted
along the $z$-direction via a Lorentz transformation with boost $\vb$. To
limit the initial violation in the constraints, $D$ is chosen to be
sufficiently large, \ie $D=240\,M_{\odot}$, and we use an optimal
composition of the two isolated-star solutions that will be presented in
a longer paper. The grid has uniform spacing $\Delta = 0.08
(0.06)\,M_{\odot}$ with extents $x/M_{\odot} \in [0, 80]$ and
$z/M_{\odot} \in [0, 150 (200)]$, where the round brackets refer to the
more demanding high-boost cases. Reflection boundary conditions are
applied on the $z=0$ plane, while outgoing conditions are used elsewhere.
A summary of the initial properties of the isolated stars leading to
supercritical models nearer to the critical line is presented in
Table~\ref{table:ID}.

\begin{figure*}
\begin{center}
\includegraphics[width=7.25cm,angle=0]{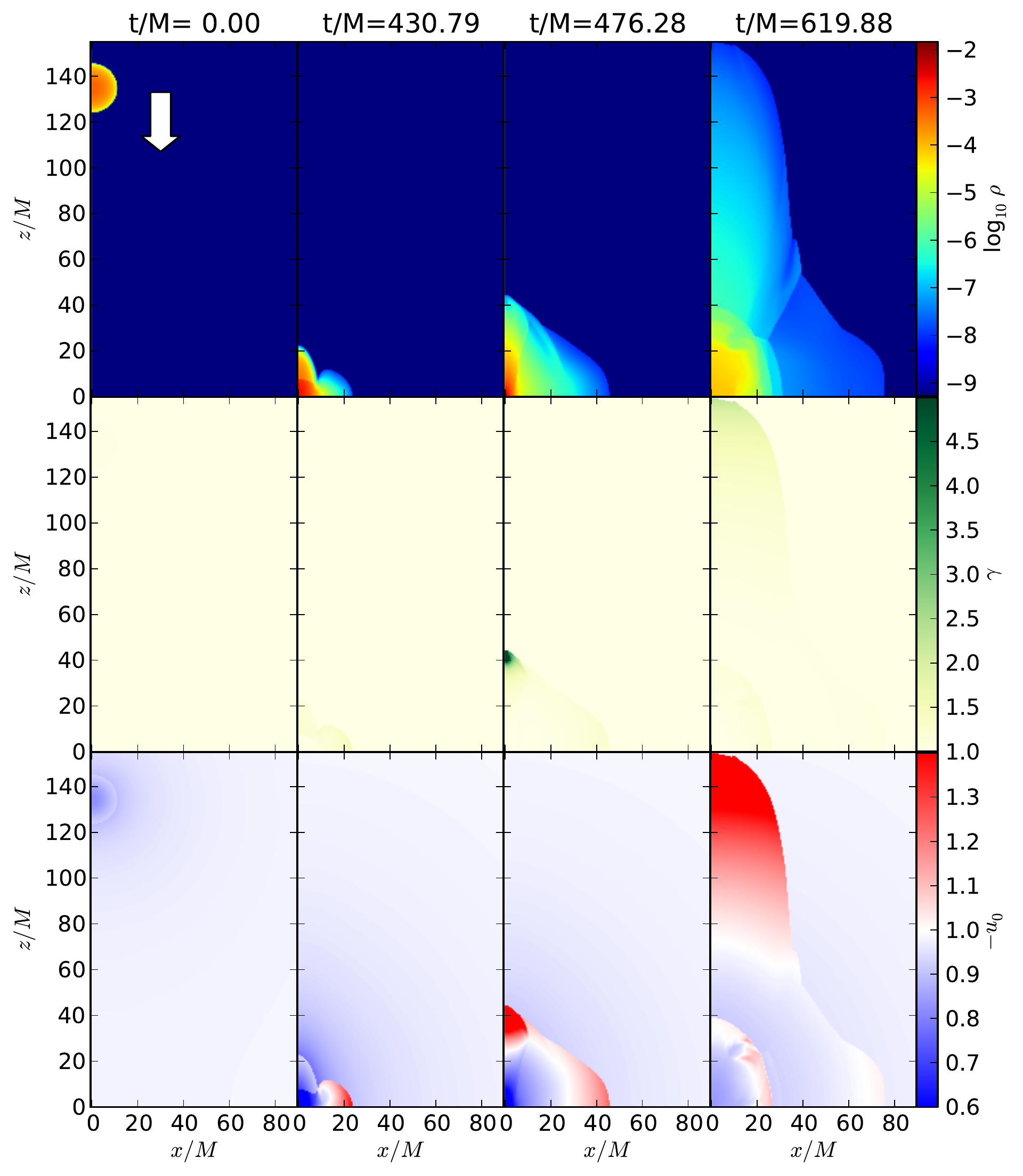}
\hskip 1.0cm
\includegraphics[width=7.25cm,angle=0]{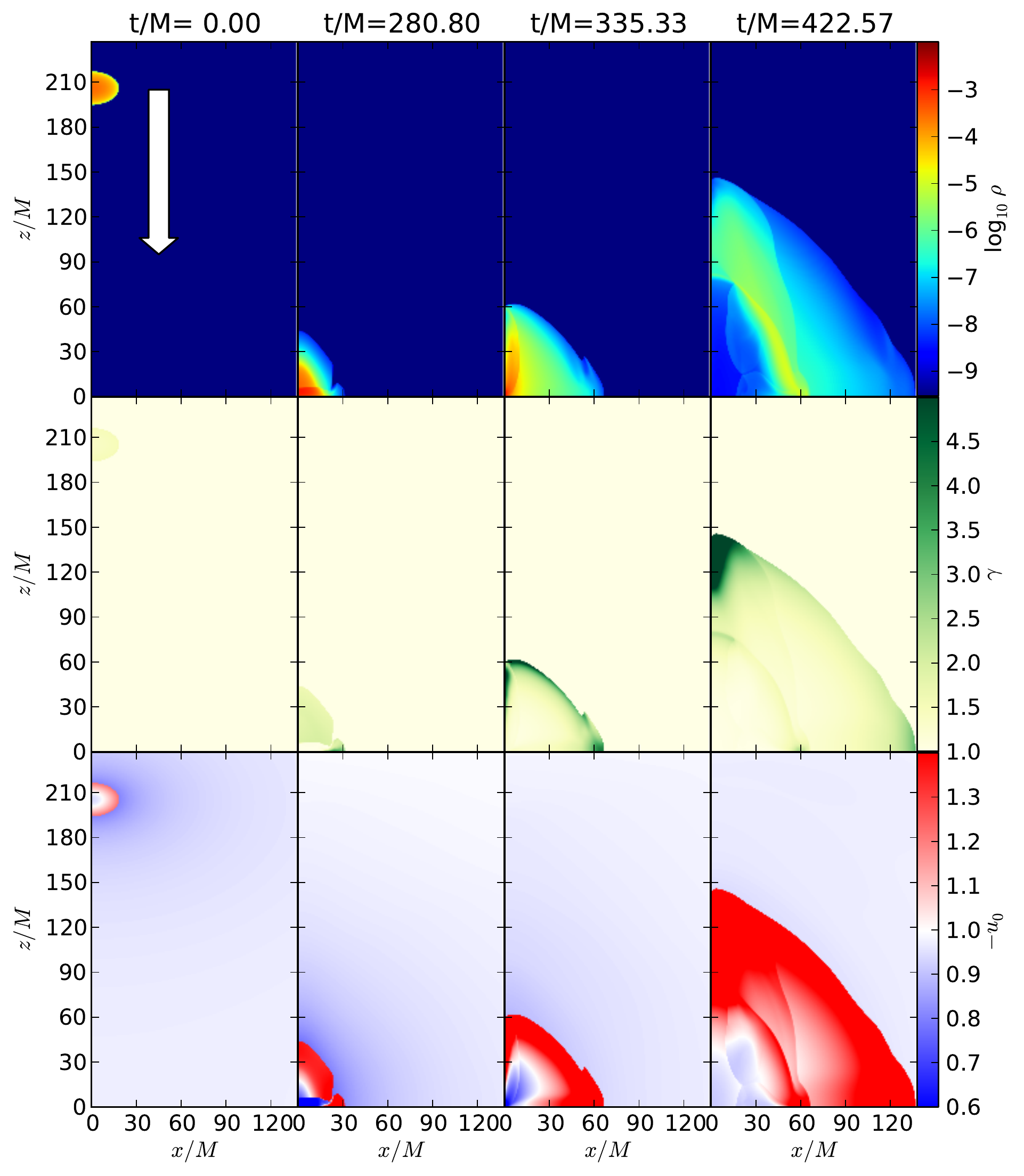}
\caption{\label{fig:fig1}Representative snapshots of the rest-mass
  density, $\rho$ in units where $c=1=M_{\odot}$ (top row), of the
  Lorentz factor, $\gamma$ (middle row), and of the local fluid energy,
  $-u_0$ (bottom row), for subcritical models with an initial small boost
  $\vb=0.3$ (left panel) or a large one $\vb=0.8$ (right panel). Note
  that the post-collision flow is essentially jet-like for the low-boost
  case (left panel), while essentially spherical for the high-boost case
  (right panel); in this latter case, most of the matter is unbound.}
  \end{center}
\end{figure*}

\section{Basic dynamics}

The dynamics of the process is rather simple. As the two stars approach
each other, the initial boost velocity increases as a result of the
gravitational attraction, leading to a strong shock as the surfaces of
the stars collide. In the case of \textit{supercritical} initial data,
\ie of stars with masses above a critical value, $M_{\rm c}$, a black
hole is promptly produced and most of the matter is accreted. Conversely,
in the case of \textit{subcritical} initial data, \ie of stars with
masses below $M_{\rm c}$, the product of the collision is a hot and
extended object with large-amplitude oscillations. Part of the stellar
matter is unbound and leaves the numerical grid as the product of the
collision reaches an equilibrium.

Figure~\ref{fig:fig1} shows snapshots at representative times of the
rest-mass density, $\rho$ (top row), of the Lorentz factor, $\gamma
\equiv (1-v^i v_i/c^2)^{-1/2}$ (middle row), and of the local fluid
energy, $-u_0$ (bottom row), for two subcritical models. The left panel,
in particular, refers to a binary boosted at $\vb=0.3$. Note that the
stars are strongly compressed by the collision, with the rest-mass
density increasing exponentially. The merged object expand in a jet-like
fashion along the $z$-direction, with the bulk of the matter being
accelerated up to $\gamma \sim 16$, or equivalently, $v/c \sim 0.998$,
but then settling on much slower flows with $\gamma \lesssim
2.1$. Furthermore, the front of the jet has $-u_0>1$ indicating that part
of the shocked matter has sufficient energy to have become
gravitationally unbound. As a result, the rest-mass density at the center
of the merged object is smaller than the maximum density of the initial
configuration, although the origin still represents the region where the
density is the largest. The right panel, on the other hand, refers to a
highly-boosted binary, \ie with $\vb=0.8$, with each star being initially
highly distorted by the Lorentz contraction. Also in this case, the stars
are strongly compressed by the collision, but the merged object expands
in a spherical blast-wave fashion, with an almost spherical distribution
of matter and bulk $\gamma$-factor. The latter reaches values as large as
$\gamma \sim 30$, or equivalently, $v/c \sim 0.999$, which, in contrast
with the low-boost case, do not decrease in time. It is worth emphasizing
that these are the first calculations exploring such ultrarelativistic
regimes in strong-curvature fields. As a comparison, the typical bulk
Lorentz factors obtained in the merger of binary neutron stars in quasi
circular orbits is $\gamma \sim 1.03$~\cite{Rezzolla:2010}. The very
large kinetic energies involved in the collision are sufficient to make a
very large portion of the stellar matter unbound, as clearly shown by the
bottom-right panel of Fig.~\ref{fig:fig1}, which reports the local fluid
energy. The rest-mass density distribution in the expanding blast wave
has a minimum at the origin, where a large rarefaction is produced by the
matter expanding as an ultrarelativistic thick shell.

The marked transition from a jet-like outflow, not too dissimilar from
the simple Bjorken flow used to model the very early states of
relativistic ion-collisions~\cite{Bjorken1983}, to a shell-like
structure, not too dissimilar from ``transverse expansion'' modelled in
the subsequent stages of relativistic ion-collisions
(see~\cite{Huovinen2006} and references therein), signals that it is not
unreasonable to extrapolate some of the results presented here also to
the collision of ultrarelativistic elementary particles.

\begin{figure}
\begin{center}
\includegraphics[width=10cm,angle=0]{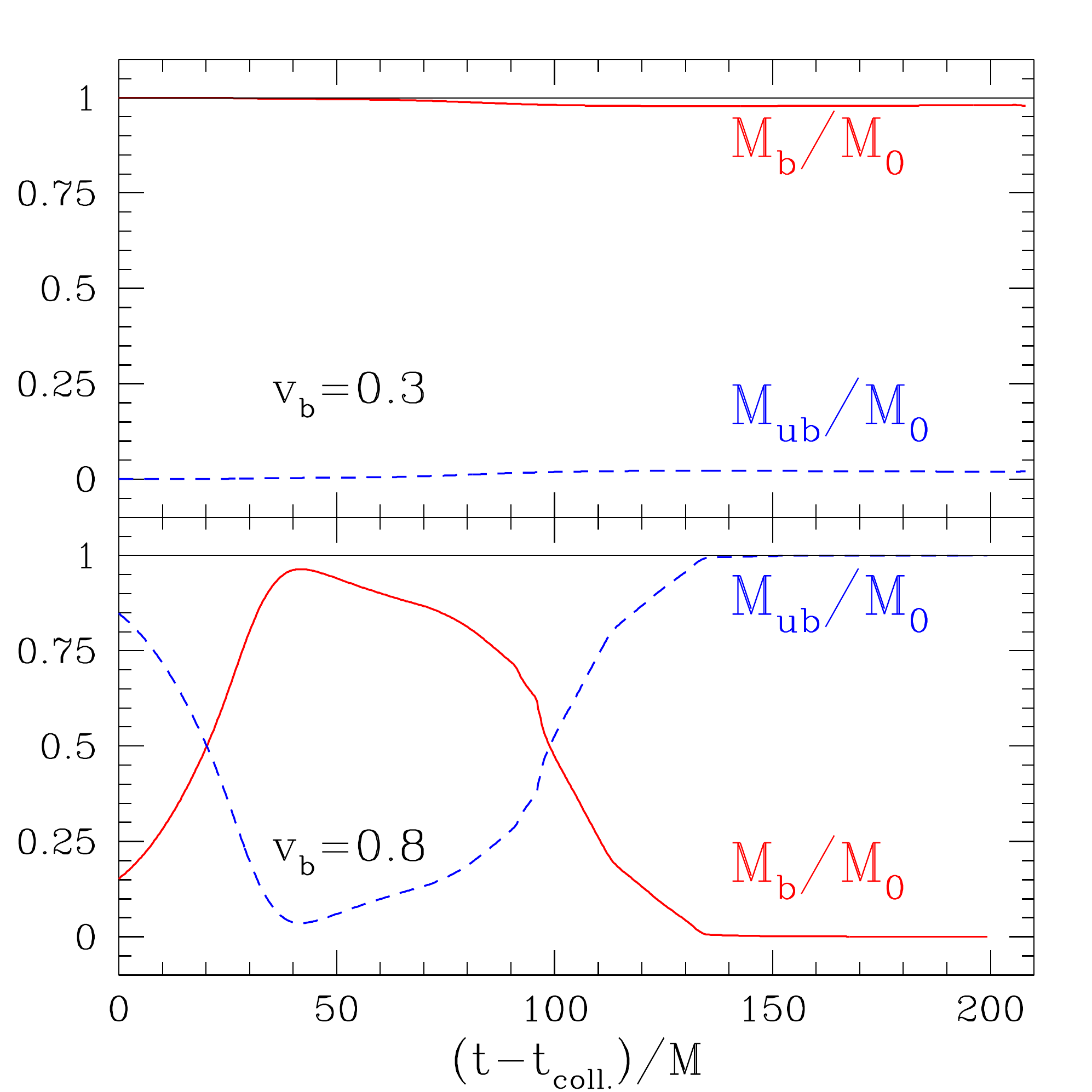}
\caption{\label{fig:fig2}Evolution in collision-retarded time of the
  fraction of the matter which is unbound, $M_{\rm ub}/M_0$ (blue dashed
  line), and the corresponding bound fraction $M_{\rm b}/M_0$ (red solid
  line), defined as the complement to the total rest-mass $M_0$ (black
  solid line).}
\end{center}
\end{figure}

The transition from the two qualitatively-different regimes discussed
above is further confirmed by the evolution of the rest-mass normalized
to the initial value $M_0$, which is shown in Fig.~\ref{fig:fig2} for the
low (top panel) and high-boost (bottom panel) cases. More specifically,
we show the fraction of the matter for which the local fluid energy $-u_0
> 1$ and which we therefore consider to be unbound, $M_{\rm ub}/M_0$
(blue dashed line). We recall that $-u_0 > 1$ is a necessary but not
sufficient condition for matter to be unbound, valid only in stationary
spacetimes and ours never really are. However, this condition is probably
also the only way to measure fraction of unbound matter. Also shown is
the corresponding bound fraction $M_{\rm b}/M_0$ (red solid line),
defined as the complement to the total rest-mass $M_0$ (black solid
line), which is conserved essentially to machine precision. Note that
this measurement is exact only in an axisymmetric and stationary
spacetime, and therefore the estimates of $M_{\rm ub}$ and $M_{\rm ub}$
are sensible only in their asymptotic values. Clearly, the unbound
fraction is just a few percent of the total rest-mass in the case of a
low-boost collision, with most of the matter being confined in the
selfgravitating ``star''. This is to be contrasted with what happens for
a high-boost collision, where the unbound fraction is $\sim 100\%$ of the
total rest-mass. This behaviour provides a strong indication that, at
least for subcritical collisions, the role played by gravitational forces
is a minor one as the kinetic energy is increased. This is what happens
in the collision of two particles at ultrarelativistic speeds, where all
of the matter is obviously unbound.

\section{Critical behaviour and scaling} 

A remarkable property of the head-on collision of compact stars is the
existence of type-I critical behaviour, which was first pointed out
in~\cite{Jin:07a} and subsequently extended in~\cite{Kellermann:08a}. In
essence, in these works it was found that when considering stars with
initial zero velocity at infinity, it is possible to fine-tune the
initial central density $\rho_{\rm c}$ (and hence the mass) near a
critical value, $\rho^{\star}_{\rm c}$, so that stars with $\rho_c >
\rho_c^{\star}$ would collapse \emph{eventually} to a black hole, while
the models with $\rho_c < \rho_c^{\star}$ would \emph{eventually} lead to
a stable stellar configuration. As a result, the head-on collision of two
neutron stars near the critical threshold can be seen as a transition in
the space of configurations from an initial stable solution over to a
critical metastable one which can either migrate to a stable solution or
collapse to a black hole~\cite{Radice:10}. As the critical limit is
approached, the survival time of the metastable object, $\tau_{\rm eq}$,
increases as $\tau_{\rm eq} = -\lambda \ln|\rho_c - \rho_c^{\star}|$,
with $\lambda \sim 10$~\cite{Jin:07a,Kellermann:08a}.

\begin{figure}
\begin{center}
\includegraphics[width=10cm,angle=0]{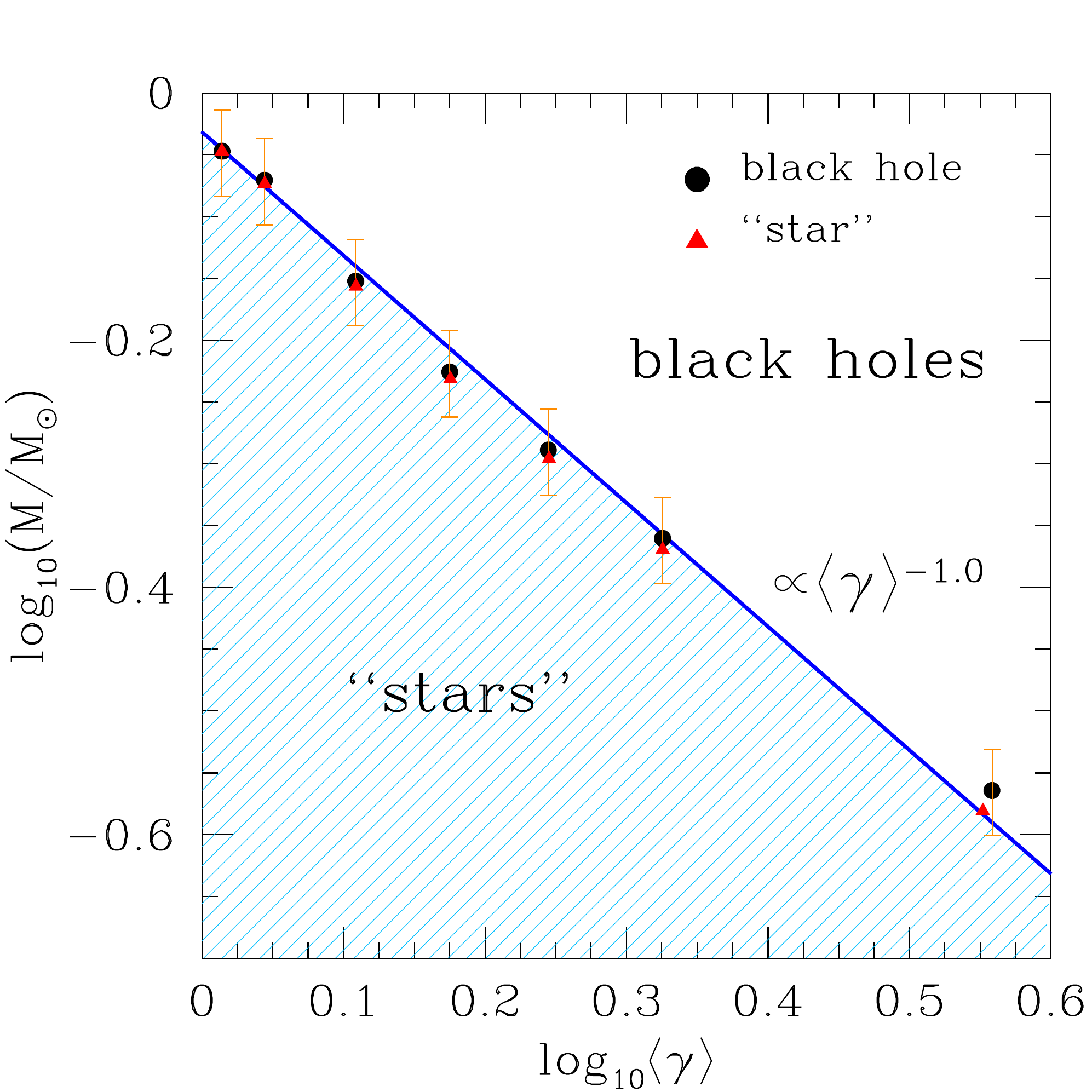}
\caption{\label{fig:fig3}Critical line as a function of the effective
  Lorentz factor, with circles indicating black holes and triangles
  selfgravitating objects.}
\end{center}
\end{figure}

Although the free-fall velocities considered
in~\cite{Jin:07a,Kellermann:08a} were very small, we find that the
critical behaviour continues to hold also when the stars are boosted to
ultrarelativistic velocities and that the threshold $\rho^{\star}_{\rm
  c}$ is a function of the initial effective boost. It should be
emphasized that determining $\rho^{\star}_{\rm c}$ becomes especially
challenging as the $\gamma$-factor is increased and the dynamics of the
matter becomes extremely violent, with very strong shocks and rarefaction
waves. However, we were able to determine the threshold for all the range
of initial boosts considered, \ie $\vb \in [0,\, 0.95]$, $\gamma_b \in
[1,\, 3.2]$, and even to a reasonable accuracy, \eg $\rho_c^{\star} = (
3.288023 \pm 0.000003 ) \times 10^{14}\,{\rm g/cm}^3$, for the initial
boost of $\vb=0.3$.
 
The existence of critical behaviour near which the details of the initial
conditions become irrelevant and which is the \emph{same} at different
boosts, \ie $\lambda$ does not depend on $\gamma$ nor on $\rho_c$
(\cite{Jin:07a,Wang2011} have shown there is ``universality'' when
varying $\gamma$ and fixing $\rho_c$), gives us a wonderful tool to
explore the conditions of black-hole formation also far away from the
masses and $\gamma$-factors considered in this paper. This is illustrated
in Fig.~\ref{fig:fig3}, which reports the gravitational mass of the
isolated spherical star as a function of the effective initial Lorentz
factor 
\begin{equation}
\label{eq:effgamma}
\langle \gamma \rangle \equiv \frac{\int d V \, T_{\mu\nu} n^{\mu} n^{\nu}}
{(\int d V T_{\mu\nu} n^{\mu} n^{\nu})_0}\,, 
\end{equation}
where $T_{\mu\nu}$ is the stress-energy tensor, $n^{\mu}$ is the unit
normal to the spatial hyperspace with proper volume element $dV$, and the
index $0$ refers to quantities measured in the initial unboosted
frame. We note that we introduce the definition of an effective Lorentz
factor~\eqref{eq:effgamma} because the stars are extended and thus the
$\gamma$-factor will be different in different parts of the
star. Expression~\eqref{eq:effgamma}, on the other hand, can be taken as
ratio of the energies measured in the boosted and unboosted frames, and
hence a generalization of the Lorentz factor for a point particle (Indeed
$\langle \gamma \rangle \to 1$ for $v_b \to 1$). Of course, other
parametrizations are possible, still leading to scaling laws, but with
slightly different exponents. An extended discussion of this point will
be presented in a longer paper. Filled circles indicate initial data
leading to a black hole, while triangles indicate initial data leading to
a ``star'', whereby we mean an object which is at least in part
selfgravitating (orange errorbars provide an approximate upper limit of
$\sim 8\%$ to the error in the measurements). Also indicated as a blue
solid line is the critical line separating the two regions of black hole
and star formation (the latter is shown as a shaded region). Clearly, the
numerical results provide a tight fit of the critical line with a power
law
\begin{equation}
\label{eq:Mth}
M_{\mathrm c}/M_{\odot} = K ~ \langle \gamma \rangle^{-n} \approx 
0.92 ~ \langle \gamma \rangle^{-1.03}
\,,
\end{equation}
and which represents the most interesting result of this work.

Expression~\eqref{eq:Mth} offers itself to a number of
considerations. First, it essentially expresses the conservation of
energy. Second, in the limit of zero initial velocities, $\langle \gamma
\rangle \to 1$, we obtain that $M_{\mathrm c} \simeq 0.92 \,M_{\odot}$,
so that the corresponding total mass, $2 M_{\mathrm c}$, is only $\sim
12\%$ larger than the maximum mass of the relative spherical-star
sequence, \ie $M_{\rm max} = 1.637\,M_{\odot}$. Third, in the opposite
limit of $\langle \gamma \rangle \to \infty$, expression~\eqref{eq:Mth}
predicts that the critical mass will go zero. This is indeed what one
would expect: as the kinetic energy diverges, no room is left for
selfgravitating matter, which will all be ejected but for an
infinitesimal amount which will go into building the zero-mass critical
black hole. Fourthly, \eqref{eq:Mth} is also in agreement with the
results in~\cite{Choptuik:2010a,East2012}, whereby one can recognize the
black-hole formation as the crossing of the critical line when moving to
larger $\gamma$-factors while keeping the rest-mass constant. Finally,
the scaling relation~\eqref{eq:Mth} can be expressed equivalently in
terms of the original stellar compactness, $M/R$ as
\begin{equation}
\label{eq:Mth_2}
\left(M/R\right)_{\mathrm c} = K' ~ \langle \gamma \rangle^{-n'} 
\approx 0.08 ~ \langle \gamma \rangle^{-1.13} \,.
\end{equation}
Since \hbox{$M_{\rm lab} \equiv \langle \gamma \rangle M$} is the mass in
the lab frame and $R$ is the largest dimension in that frame being the
transverse one to the motion, the ratio $\left(M_{\rm
  lab}/R\right)_{\mathrm c} = K' \langle \gamma \rangle^{1-n'}\sim K'
\langle \gamma \rangle^{-0.13}$ provides the condition for the amount of
energy that, when confined in a hoop of radius $R$, would lead to a black
hole, as in the spirit of the hoop conjecture.

\section{Conclusions}

We have computed the first ultrarelativistic collisions of compact fluid
stars leading to black-hole formation. The Lorentz factor reached in our
relativistic-hydrodynamic simulations in strong-field regimes are
considerably larger than those ever explored in inspiralling neutron-star
binaries. We find that the properties of the flow after the collision
change with Lorentz factor, with most of the matter being ejected in a
spherical blast wave for large boosts. As in previous work, we find that
the collision exhibits a critical behaviour of type I and that this
persists also as the initial boost is increased. This allows us to use a
simple scaling law with a transparent physical interpretation and a
condition for black-hole formation in the spirit of the hoop conjecture.

\section{Acknowledgements}

We thank J. L. Jaramillo, D. Radice, J. Miller, A. Harte, W. East, and
F. Pretorius for useful discussions. This work was supported in part by
the DFG grant SFB/Transregio~7 and by ``CompStar'', a Research Networking
Programme of the ESF. KT is supported by a JSPS Postdoctoral Fellowship
for Research Abroad. All simulations were performed on clusters at the
AEI.

\clearpage

\section*{References}
\bibliographystyle{unsrt.bst}

\end{document}